\pdfoutput=1

\documentclass[11pt,letter]{article}

\usepackage{graphicx}
\usepackage{tabularx}
\usepackage{amsopn}
\usepackage[cmex10]{amsmath}
\usepackage{algorithm}
\usepackage{algorithmic}
\usepackage[algo2e,ruled,linesnumbered]{algorithm2e}
\usepackage{bm}
\usepackage{booktabs}
\usepackage{epsfig}
\usepackage{subfigure}
\usepackage{amsfonts}
\usepackage{comment}
\usepackage{enumerate}
\usepackage{mathrsfs} 
\usepackage{flushend}
\usepackage[perpage,symbol]{footmisc}   \setfnsymbol{wiley}   
\usepackage{colortbl} 
\usepackage{url}
\usepackage{authblk}
\usepackage[pdfborder={0 0 0}]{hyperref}

\DeclareMathOperator*{\argmax}{arg\,max}

\begin{document}
\title{A framework for community detection in heterogeneous multi-relational networks}

\author{Xin Liu\,$^{1,2,3}$,
    Weichu Liu\,$^1$,
    Tsuyoshi Murata\,$^1$,
    and Ken Wakita\,$^{1,2}$}

\affil{$^1$Tokyo Institute of Technology\\
    2-12-1 Ookayama, Meguro, Tokyo, 152-8552 Japan}

\affil{$^2$CREST JST\\
    K's Gobancho, 7, Gobancho, Chiyoda, Tokyo, 102-0076 Japan}

\affil{$^3$Wuhan University of Technology\\
    122 Luoshi Road, Wuhan, Hubei, 430070 China}

\affil{E-mail: \{tsinllew@ai.cs, liu.w@ai.cs, murata@cs, wakita@is\}.titech.ac.jp}

\maketitle

\begin{abstract}
There has been a surge of interest in community detection in homogeneous single-relational networks which contain only one type of nodes and edges. However, many real-world systems are naturally described as heterogeneous multi-relational networks which contain multiple types of nodes and edges. In this paper, we propose a new method for detecting communities in such networks. Our method is based on optimizing the composite modularity, which is a new modularity proposed for evaluating partitions of a heterogeneous multi-relational network into communities. Our method is parameter-free, scalable, and suitable for various networks with general structure. We demonstrate that it outperforms the state-of-the-art techniques in detecting pre-planted communities in synthetic networks. Applied to a real-world Digg network, it successfully detects meaningful communities.
\end{abstract}

\maketitle

\section{Introduction}\label{sec1}
Many social, biological, and information systems can be described as networks, where nodes represent fundamental entities of the system, such as individuals, users, genes, web pages, and so on; and edges represent relations or interactions between the entities. In recent years, there has been a surge of interest in analysis of networks. A highly discussed topic is community detection \cite{GirvanNewmanCommuityDefinition} --- the identification of groups of closely related nodes that correspond to functional subunits of the underlying system, such as collections of pages on closely related topics on the web or groups of people with common interests in social media. Thus community detection provides insight into how the system is internally organized.

\begin{figure}[!t]
\centering
\includegraphics[width=0.60\textwidth]{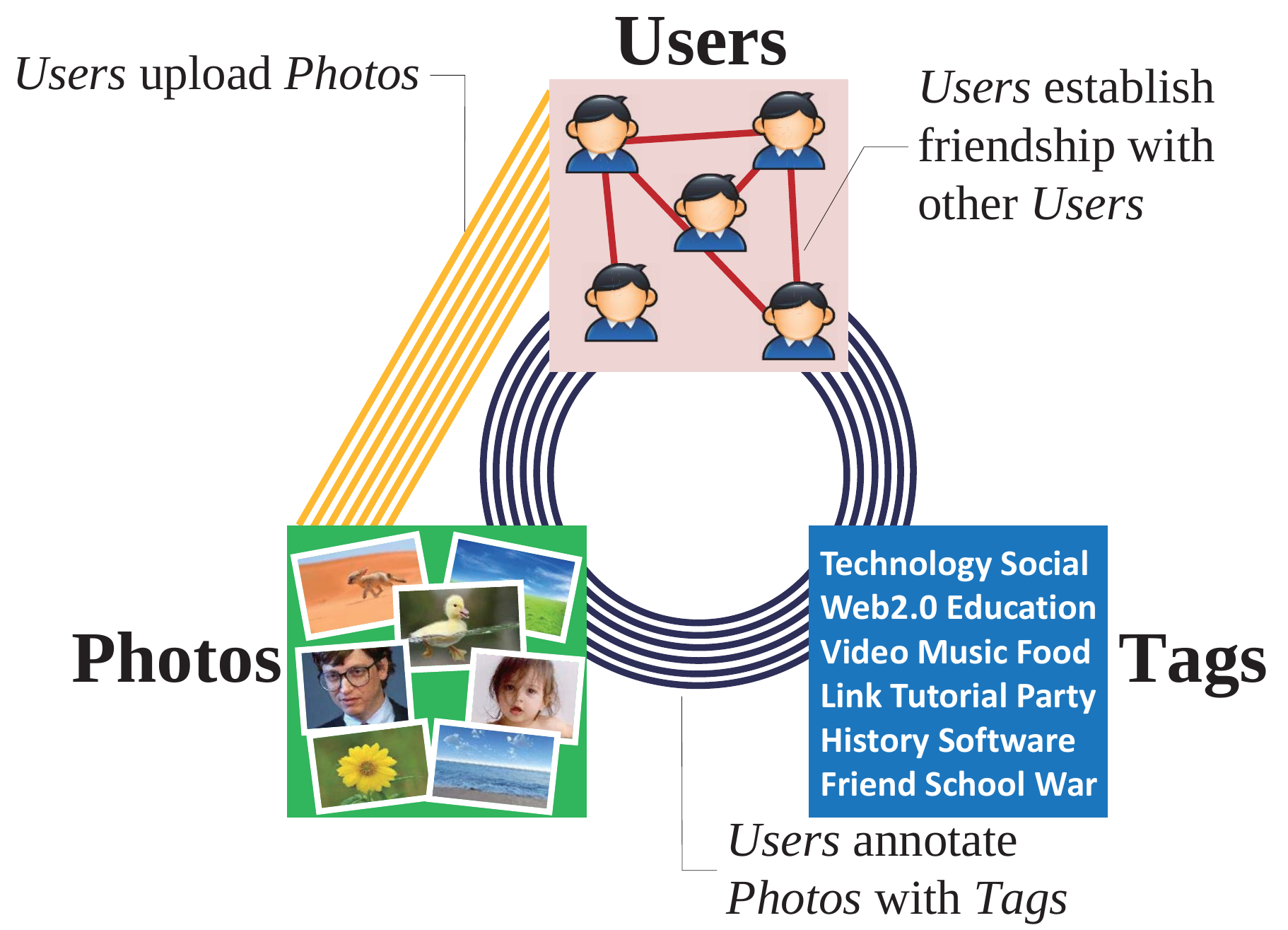}
\caption{\label{fig1} Describing Flickr as a heterogeneous multi-relational network. This network contains three types of nodes: 1) users, 2) photos, 3) tags; and three types of edges: 1) the edges representing friendship between users, 2) the edges representing that users unload photos, 3) the 3-way hyper-edges representing that users annotate photos with tags.}
\end{figure}

Previous studies on community detection mainly focus on homogeneous single-relational networks which contain only one type of nodes and edges. Many real-world systems, however, are naturally described as \textit{heterogeneous} (contain multiple types of nodes) \textit{multi-relational} (contain multiple types of edges) networks. Take the photo service website Flickr as an example. Flickr users can upload photos, annotate photos with tags, and establish friendship with other users. As shown in Fig.~\ref{fig1}, Flickr can be described as a heterogeneous multi-relational network, which contains three types of nodes (users, photos, and tags) and three types of edges representing the above relations.

Traditionally we first simplify a heterogeneous multi-relational network to a homogeneous single-relational network and then conduct community analysis. Take the above Flickr network as an example. We can omit information about photos and tags, and detect communities directly in the homogeneous user friendship network. However, we may omit too much valuable information in this simplification process. For example, the data about users can be incomplete and noisy, and some active users have thousands of friends while some have no friends at all. Consequently, we cannot obtain real user communities based on the friendship network alone. In this scenario, we expect that a method which effectively utilizes multi-faceted information in the heterogeneous multi-relational network would enhance the community detection results.

In this paper, we propose a new method for detecting communities in heterogeneous multi-relational networks. Our method follows the line of the modularity optimization method which is widely used for detecting communities in homogeneous single-relational networks \cite{NewmanGreedy,ClausetFastGreedy,NewmanModularityEigenvectors,BlondelFastUnfolding,DuchExtremalOptimization,MedusSA,WakitaCommunityMegaScaleNetwork,SchuetzMultistepGreedy,SchuetzMultistepGreedy2,BarberLPAm,LiuLPAmplus,ShiModularityGA,HeModularityAntColonyOpt}. In specific, we first propose a \textit{composite modularity} for evaluating the ``goodness'' of a partition of a heterogeneous multi-relational network into communities. Then we develop a fast algorithm for optimizing the composite modularity and detecting the communities. Our method is consistent with the modularity optimization method, because the composite modularity reduces to modularity when we deal with a homogeneous single-relation network. Our method is applicable to networks of general structure, which may even contain hyper-edges for representing relations between more than two nodes. Another advantage of our method is that it is parameter-free and can automatically detect communities without any a priori knowledge such as the number of communities. Experiments in synthetic networks show that our method outperforms the state-of-the-art techniques in detecting pre-planted communities. We use a real-world Digg network to illustrate that our method successfully detects meaningful communities.

The rest of the paper is organized as follows. Section~\ref{sec2} reviews related research. Section~\ref{sec3} formulates the problem of community detection in a general heterogeneous multi-relational network. Section~\ref{sec4} introduces our new method. Section~\ref{sec5} presents experimental results, followed by a conclusion in Section~\ref{sec6}.

\begin{figure}[!t]
\centering
\setlength{\abovecaptionskip}{1.5pt}
\subfigure[][]{\label{fig2a}
\includegraphics[width=0.24\textwidth]{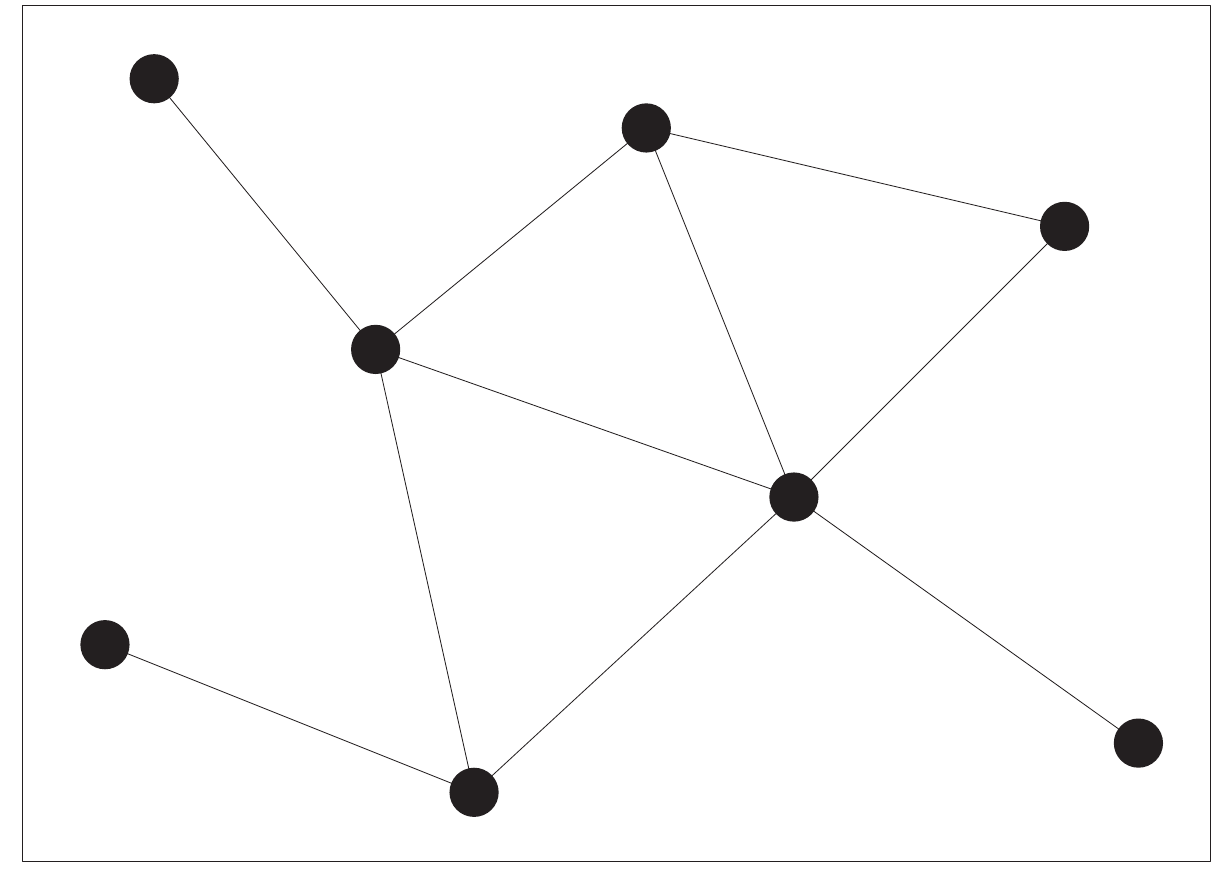}
}\qquad
\subfigure[][]{\label{fig2b}
\includegraphics[width=0.24\textwidth]{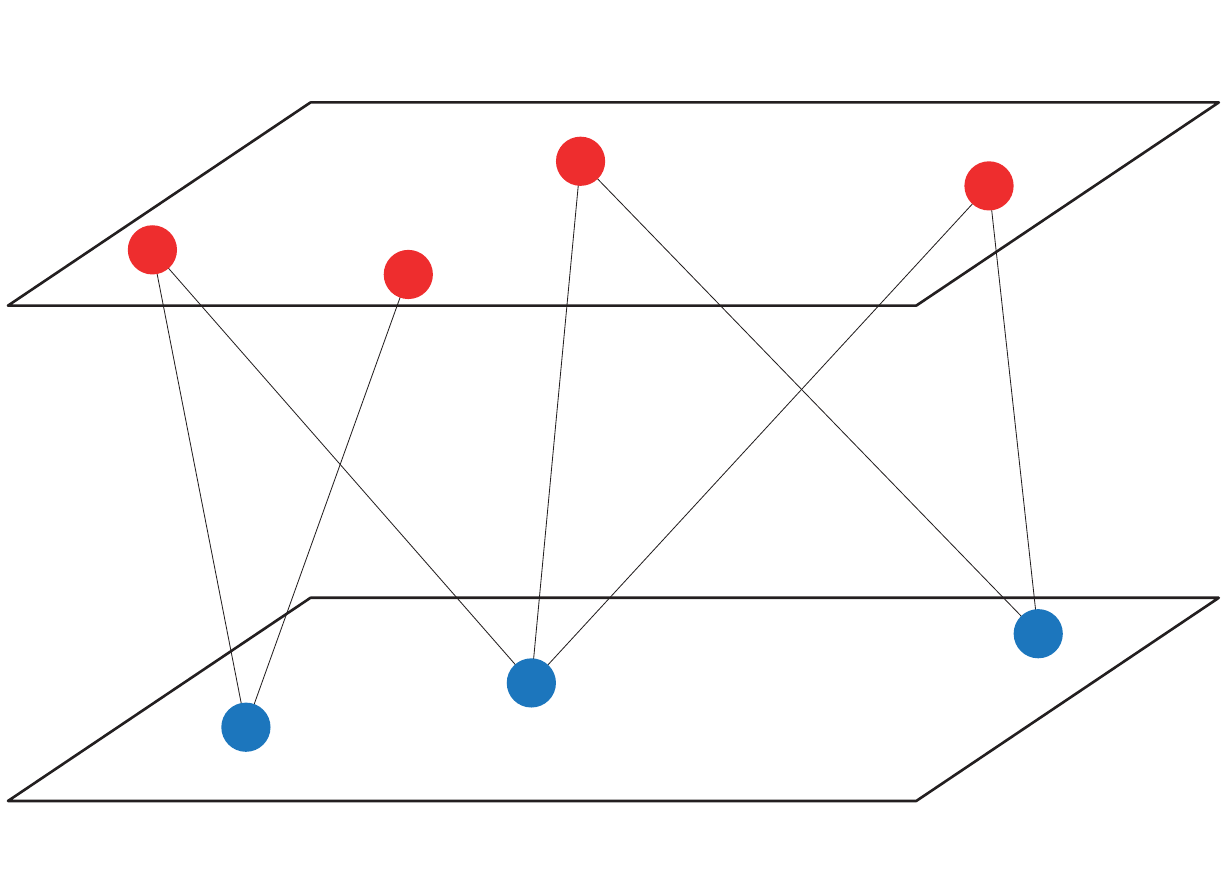}
}\qquad
\subfigure[][]{\label{fig2c}
\includegraphics[width=0.24\textwidth]{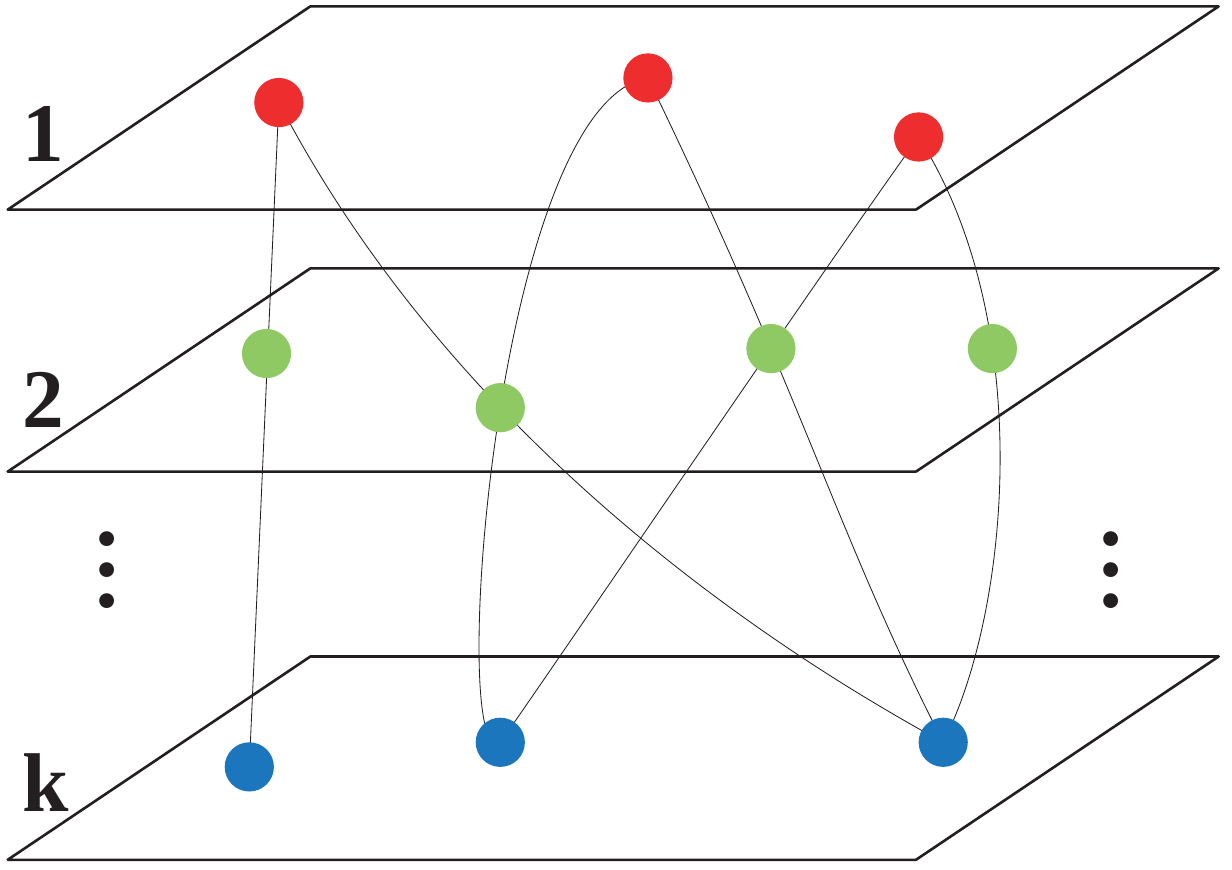}
}
\caption{\label{fig2} (a) A unipartite network. (b) A bipartite network. (c) A k-partite network.}
\end{figure}

\section{Related Work}\label{sec2}
The study of community detection in homogeneous single-relational networks or called \textit{unipartite networks} (see Fig.~\ref{fig2a}) has a long history. It is closely related to graph partitioning \cite{KernighanGraphPartitioning} in computer science, and hierarchical clustering \cite{ScottSocialNetworkAnalysis} in sociology. In the past decade, this study has attracted a great deal of interest and various methods were proposed \cite{FortunatoCommunityDetectionReview,NewmanCommunityReview,DanonCompareCommunityAlgo,LancichinettiCommunityAlgoAnalysis,OrmanComparativeEvaluationCommunityDetectionAlg}. In particular, a family of methods which are widely used is known as \textit{modularity optimization}. Modularity was originally proposed by Newman and Girvan \cite{GirvanNewmanCommuityDefinition} for evaluating the ``goodness'' of a partition of a unipartite network into communities. The definition of modularity involves a comparison of the fraction of intra-community edges in the observed network minus the expected value of that fraction in a randomized network, which is called the null model. More precisely, the mathematical expression of modularity in an undirected single-relational network reads
\begin{eqnarray}
\mathrm{Q}(\mathcal{L})=\frac{1}{2m} \sum_{i=1}^{n}\sum_{j=1}^{n} (A_{ij}-P_{ij})\,\delta(l_{i},l_{j}), \label{eq1}
\end{eqnarray}
where $n$ is the number of nodes, $m$ is the number of edges, $\mathcal{L}=\{l_1,l_2,\cdots,l_n\}$ is a partition, with element $l_{i}$ indicating the community membership of the $i$-th node $v_i$, ${A}_{ij}$ is the number of edges between $v_i$ and $v_j$ in the observed network, ${P}_{ij}$ is the expected value of that number in the null model, and $\delta$ is the Kronecker's delta. In addition, another equivalent expression of modularity is
\begin{eqnarray}
Q(\mathcal{L}) = \sum_{{\alpha}=1}^{c} (e_{\alpha\alpha}-a_{\alpha}^2), \label{eq2}
\end{eqnarray}
where
\begin{eqnarray}
e_{\alpha\alpha} = \frac{1}{2m} \sum_{i=1}^{n}\sum_{j=1}^{n} A_{ij} \delta(l_{i},\alpha)\delta(l_{j},\alpha) \label{eq3}
\end{eqnarray}
is the fraction of edges within community $V_{\alpha}$, and
\begin{eqnarray}
a_{\alpha} = \frac{1}{2m} \sum_{i=1}^{n}\sum_{j=1}^{n} A_{ij} \delta(l_{i},\alpha) \label{eq4}
\end{eqnarray}
is the fraction of edges attached to community $V_{\alpha}$.

Optimizing modularity is proved to be NP-hard \cite{BrandesModularityNPCompleteness}. Researchers have developed various heuristic optimization algorithms \cite{NewmanGreedy,ClausetFastGreedy,NewmanModularityEigenvectors,BlondelFastUnfolding,DuchExtremalOptimization,MedusSA,WakitaCommunityMegaScaleNetwork,SchuetzMultistepGreedy,SchuetzMultistepGreedy2,BarberLPAm,LiuLPAmplus,ShiModularityGA,HeModularityAntColonyOpt}. In particular, the simulated annealing algorithm \cite{MedusSA} is the most accurate (in terms of the modularity score) \cite{DanonCompareCommunityAlgo}. However, this algorithm requires a long time to complete, and is only suitable for small-scale networks. On the other hand, the label propagation algorithm \cite{BarberLPAm}, which requires only near linear time to complete, is perhaps the fastest. However, this algorithm tends to get stuck in a poor local optimum \cite{LiuLPAmplus}. In practice, the Louvain algorithm \cite{BlondelFastUnfolding} is widely used, since it reaches a proper balance between accuracy and speed.

A notable issue of modularity optimization is the resolution limit, which refers to the incapability of detecting small communities in large-scale networks \cite{FortunatoResolutionLimit,GoodModularityDegeneracy}. Researchers have tried to get around this issue by proposing variants of modularity. For example, Arenas et al. modified modularity by adding a parameter that forms self-loop for each node \cite{ArenasDifferentResolutionLevels}. Reichardt and Bornholdt modified modularity by adding a parameter in front of the null model term \cite{ReichardtStatisticalMechanicsCommunityDetection}. Both parameters can be used to control the resolution level and detect communities at multiple resolutions. However, a recent study by Lancichinetti and Fortunato demonstrated that these methods are intrinsically deficient and still suffer from the resolution limit \cite{lancichinettiModularityLimit}.

There are studies on community detection in heterogeneous single-relational networks. The most simple model of such networks is the bipartite network, where there are two types of nodes and edges that connect nodes of different types (see Fig.~\ref{fig2b}). Examples of bipartite networks include author-paper networks, actor-movie networks, and customer-product networks. A direct generalization of bipartite network is the k-partite network, where there are k types of nodes and hyper-edges that connect k nodes of different types (see Fig.~\ref{fig2c}). Researchers extended modularity to bipartite networks and k-partite networks. For example, Guimer\`a et al. proposed a bipartite modularity which focuses on evaluating the partition of only one type of nodes, and used simulated annealing algorithm for optimization \cite{GuimeraBipartiteModularity}. Barber proposed a bipartite modularity which assumes a bipartite structure of the null model, and devised a label propagation algorithm called BRIM \cite{BarberBipartiteModularity} for optimization. Murata proposed a k-partite modularity in a unified way as the definition of modularity, in the sense that his k-partite modularity can reduce to modularity if the k-partite network becomes a unipartite network \cite{MurataBipartiteModularity,MurataTriModularityWWW}. Neubauer et al. proposed another k-partite modularity by reducing a k-partite network to $\binom{\text{k}}{2}$ bipartite networks and utilizing Murata's definition of bipartite modularity \cite{NeubauerKPartiteCommunity}. Murata and Neubauer et al. employed greedy bottom-up algorithm for optimization. In addition to the family of methods based on optimizing variants of modularity, Sun et al. and Liu et al., respectively, proposed information compression based methods for bipartite networks \cite{SunGraphscopeMiningTimeEvolvingGraph} and k-partite networks \cite{LiuCommunityDetectionKKNetworkJCST}. Moreover, there are methods for simultaneously clustering related sets of heterogeneous data, such as documents and words. Such methods are often referred to co-clustering \cite{ChoCoclusteringGeneExpressionData,DhillonInfoCoClustering,LongCoclusteringBlockValueDecomposition}.

There are studies on community detection in homogeneous multi-relational networks (sometimes called the multi-mode networks, multi-dimensional networks, or multi-slice networks). For example, researchers developed methods for detecting communities in a particular subclass of such networks, known as signed networks where each edge has a positive or negative sign \cite{TraagCommunityDetectionNetworkPositiveNegativeLinks,YangCommunitySignedNetwork,BogdanovCommunitySignedNetwork}. Mucha et al. proposed a multiplex model for describing a homogeneous multi-relational network and developed a method based on optimizing a generalized modularity known as stability \cite{MuchaMultiplexNetwork}. Moreover, researchers proposed methods based on matrix approximation \cite{TangIdentifyEvolvingGroupsMultimodeNet} and spectral analysis \cite{TangCommunityDetectionHeterogeneousInteractionAnalysis}.

Recent research have also addressed the problem of community detection in heterogeneous multi-relational networks. Comar et at. developed a method based on non-negative matrix factorization \cite{ComarCommunityDetectionMultipleNetwork}. However, their work is restricted to a specific subclass of networks which contains two types of nodes and three types of edges. Sun et al. designed a ranking-based community detection method for a specific subclass of networks, referred to the star network schema \cite{SunClusteringStarHeterogeneousNetwork}. Thus both of these two methods are not applicable to general networks with any possible structure. As for the problem in a general heterogeneous multi-relational network, a naive approach is to simplify the network to a single-relational network and then conduct community detection. However, valuable information might be omitted, leading to inaccurate results. Popescul et al. proposed a method by calculating the similarity between nodes and building a similarity matrix \cite{PopesculClusteringBasedOnSimilarity}. However, when the structure of a network becomes complex, it is difficult to find a reasonable similarity measure. Also, high computational complexity is another issue, which prevents this method from being applied to large-scale networks. In addition, Lin et al. proposed a method based on tensor factorization \cite{LinMetaFac}. Different from our work, they suppose that a community contains nodes of different types. If we partition nodes of different types separately, it means that nodes of different types have the same number of communities. Unfortunately, this situation is rarely seen in real-world scenario. Another important work is Ref.\,\cite{SunClusteringHeterogeneousInfoNetworkAttribute}, where the authors considered community detection in networks with incomplete attributes. The difference from our work is that their method utilizes both edges and node attributes. Note that a drawback of the above methods is that they require a priori knowledge about the number of communities. This limits their usage in inferring the latent organization of a real system.

\begin{figure}[!t]
\centering
\includegraphics[width=0.6\textwidth]{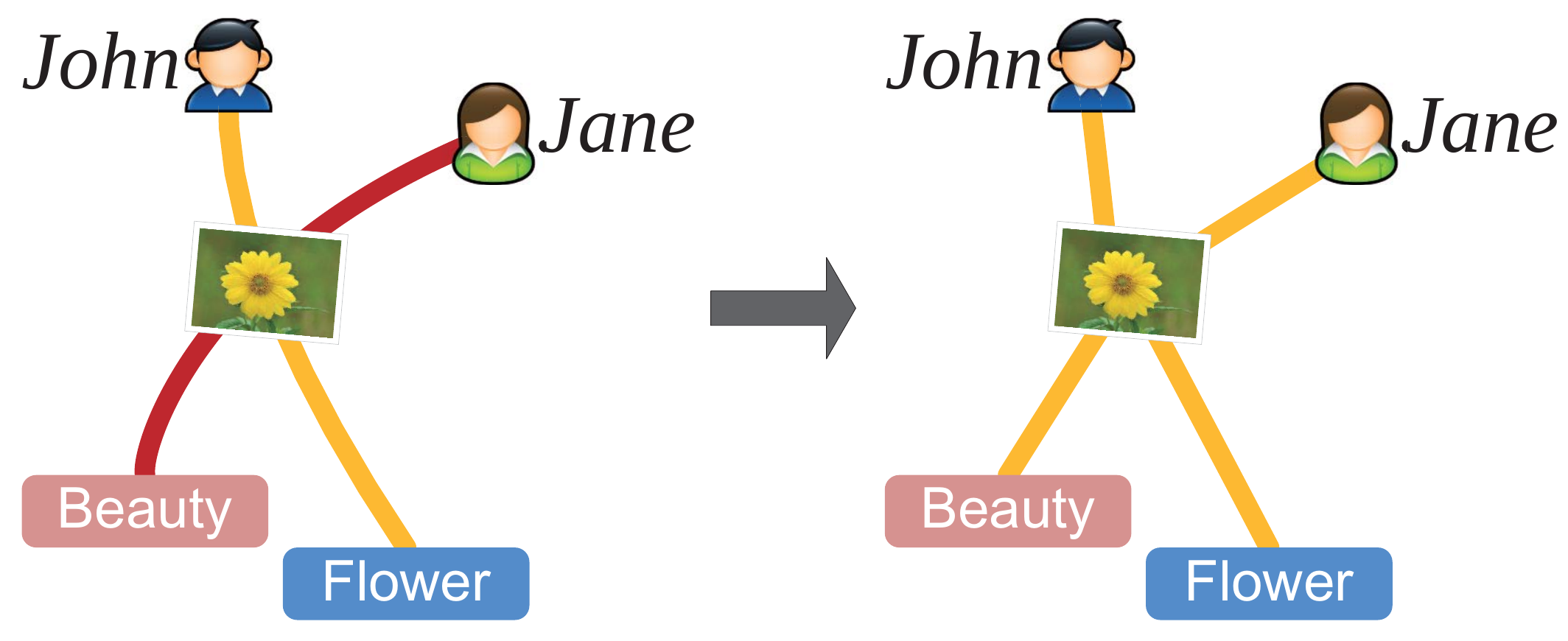}
\caption{\label{fig3} Reducing hyper-edges to normal edges causes information loss. The curved lines indicate 3-way hyper-edges.}
\end{figure}

Hyper-edges are seldom discussed in previous studies. In general, hyper-edges are useful for representing relations that involve more than two nodes. For example, in a social tagging system such as Flickr, users use tags to annotate photos. Such a relation can be naturally represented as a 3-way hyper-edge. Although we can reduce the 3-way hyper-edge to a normal edge, some information is lost during the reduction process. As shown in Fig.~\ref{fig3}, \textit{John} annotates the photo with \textit{Flower}, and \textit{Jane} annotates the photo with \textit{Beauty}. If we reduce the hyper-edges to normal edges, we cannot distinguish who uses \textit{Flower} and who uses \textit{Beauty}. In this paper, we aim at automatically detecting communities in a general heterogeneous multi-relational network which may contain hyper-edges, without a priori knowledge about the number of communities.

\begin{figure}[!t]
\setlength{\abovecaptionskip}{-7pt} 
\begin{center}
\subfigure[]{\label{fig4a}
\begin{minipage}[t]{0.4\textwidth}
\includegraphics[width=1\textwidth]{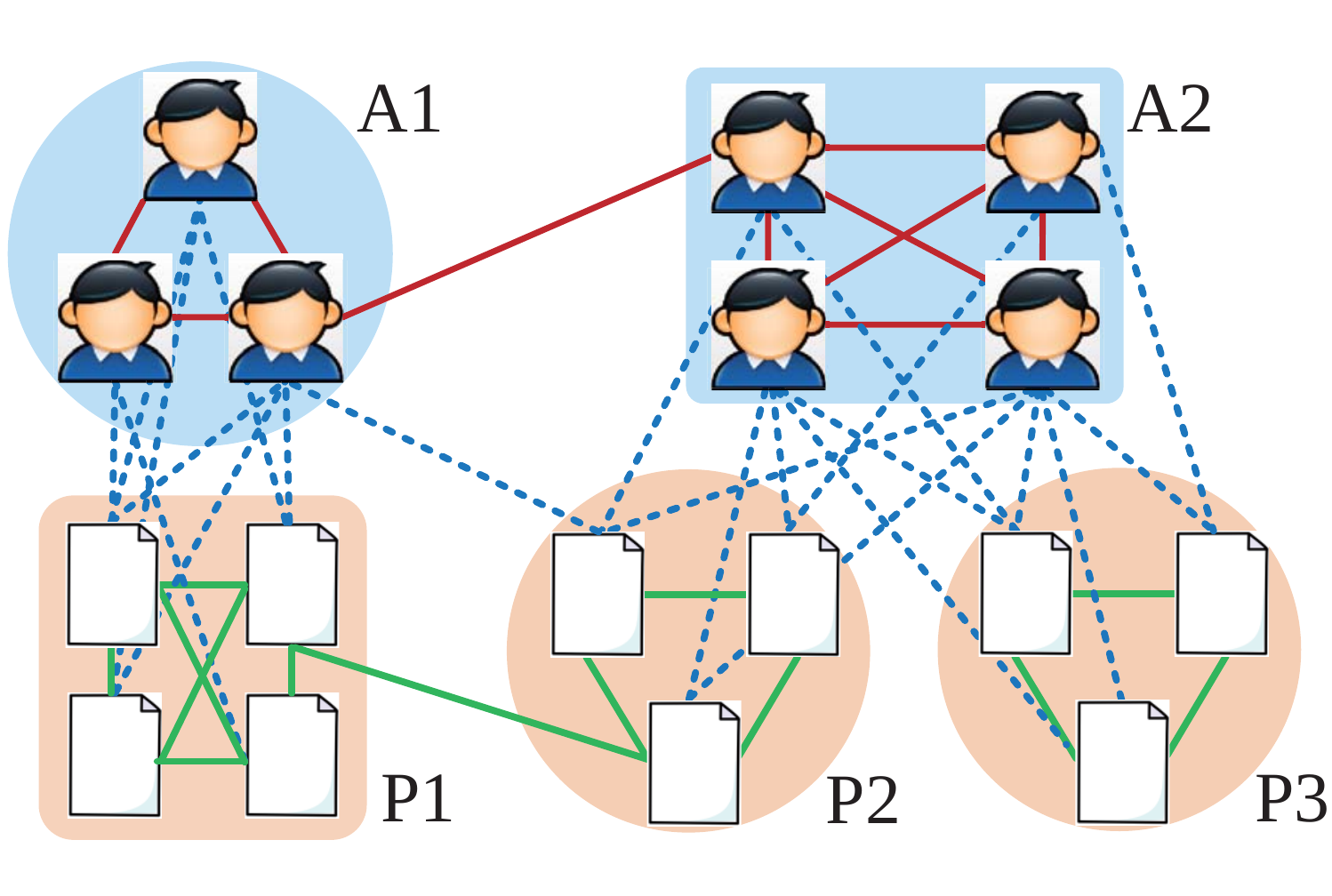}
\end{minipage}
}\qquad
\subfigure[]{\label{fig4b}
\begin{minipage}[t]{0.4\textwidth}
\includegraphics[width=1\textwidth]{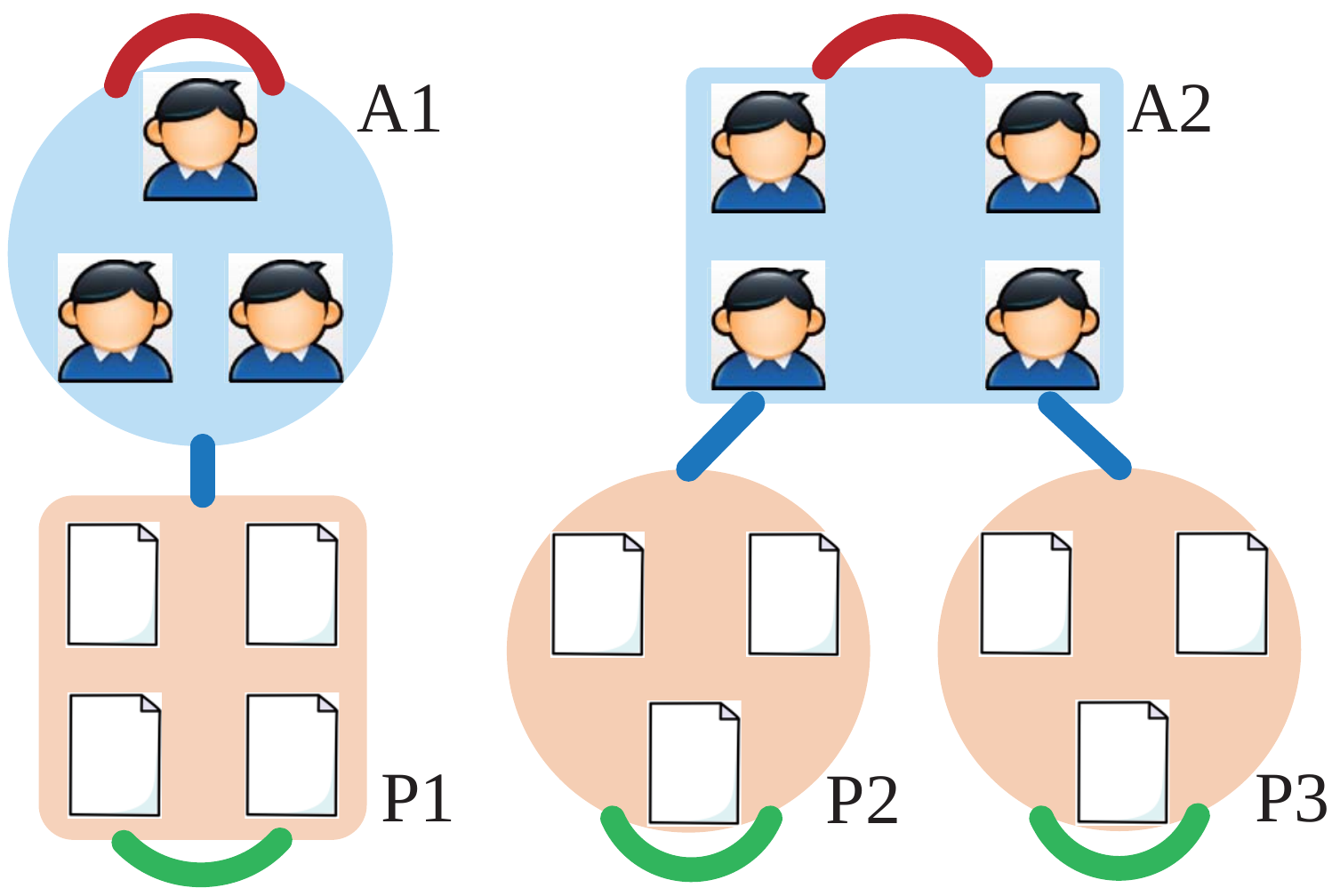}
\end{minipage}
}
\end{center}
\caption{\label{fig4} (a) The link pattern based community. (b) The link patterns of the communities.}
\end{figure}

\section{Problem Formulation}\label{sec3}
In homogeneous single-relational networks, people mainly focus on densely intra-connected and sparsely inter-connected community. In heterogeneous multi-relational networks, this concept can be generalized to the link pattern based community \cite{LongLinkPattern,LongGraphClustering} --- a group of nodes that have the similar link patterns, i.e., the nodes within a community connect to other nodes in similar ways. Fig.~\ref{fig4a} shows a heterogeneous multi-relational network with two types of nodes (author and paper nodes), and three types of edges (the edges representing the friendship between authors, the authorship between authors and papers, and the citation relationship between papers). This network has two author communities (A1 and A2), and three paper communities (P1, P2, and P3). Take the community A1 as an example. The nodes in A1 have the similar link patterns, as they all densely connect to the nodes in A1 and P1, and sparsely connect to the nodes in A2, P2, and P3. Similar interpretation applies to other communities. Fig.~\ref{fig4b} shows the link patterns of these communities. Note that the definition of link pattern based community is reasonable, as the nodes with similar link patterns are likely to share common features and form the real community.

In the following, we formulate the problem of community detection in a general heterogeneous multi-relational network. Now suppose a heterogeneous multi-relational network $\mathbf{G} = (\mathbf{V}^{[1]} \cup \mathbf{V}^{[2]} \cup \cdot\!\cdot\!\cdot \cup \mathbf{V}^{[r]},\,\mathbf{E}^{[1]} \cup \mathbf{E}^{[2]} \cup \cdot\!\cdot\!\cdot \cup \mathbf{E}^{[s]})$, where there are $r$ types of nodes and $s$ types of edges. $\mathbf{V}^{[x]}$ is the node set of the $x$-th type. $\mathbf{E}^{[y]}$ is the edge set of the $y$-th type. $\mathbf{E}^{[y]}$ should satisfy either of the following two conditions:
\begin{enumerate}
\item There exists a $x \in \{1,2,\cdots,r\}$, such that $\mathbf{E}^{[y]} \subseteq \mathbf{V}^{[x]} \times \mathbf{V}^{[x]}$, i.e., $\mathbf{E}^{[y]}$ is a set of edges that connect nodes of the same type.
\item There exists $x_1,x_2,\cdots,x_k \in \{1,2,\cdots,r\}$ ($k \le r$) which are not equal to each other, such that $\mathbf{E}^{[y]} \subseteq \mathbf{V}^{[x_1]} \times \mathbf{V}^{[x_2]} \times\cdots\times \mathbf{V}^{[x_k]}$, i.e., $\mathbf{E}^{[y]}$ is a set of $k$-way edges \footnote{If $k>2$ the edges are actually hyper-edges.} that connect nodes of different types.
\end{enumerate}

Given $\mathbf{G}$, the problem is to find a ``good'' partition $\mathcal{L}=\mathcal{L}^{[1]}\cup\mathcal{L}^{[2]}\cup\cdots\cup\mathcal{L}^{[r]}$, such that $\mathcal{L}^{[x]}$ divides $\mathbf{V}^{[x]}$ into disjoint communities $(x=1,\cdot\!\cdot\!\cdot,r)$. The meaning of ``good'' is that the nodes in each community have the similar link patterns. Note that the number of communities in each node set is not known a priori.

\begin{table}[!t]
\setlength{\belowcaptionskip}{1pt}
\centering    
\caption{\label{table1}Notations for a heterogeneous multi-relational network $\mathbf{G}$}
\fontsize{10pt}{8pt}\selectfont
\begin{tabularx}{0.97\textwidth}{p{0.001\textwidth}lp{0.03\textwidth}Xp{0.001\textwidth}}
\toprule
& \textrm{Symbol}                                         && \textrm{Meaning} &\\
\midrule
& $n$                                                     && The total number of nodes &\\
& $m$                                                     && The total number of edges &\\
& $r$                                                     && The number of node types &\\
& $s$                                                     && The number of edge types &\\
& $\mathbf{V}^{[x]}$                                      && The node set of the $x$-th type &\\
& $\mathbf{E}^{[y]}$                                      && The edge set of the $y$-th type &\\
& $\mathbf{G}^{[y]}$                                      && The subnetwork consisting of $\mathbf{E}^{[y]}$ and the incident nodes&\\
& $\mathbf{A}^{[y]}$                                      && The connectivity array of $\mathbf{G}^{[y]}$&\\
& $n^{[x]}$                                               && The number of nodes in $\mathbf{V}^{[x]}$ &\\
& $c^{[x]}$                                               && The number of communities in $\mathbf{V}^{[x]}$ &\\
& $m^{[y]}$                                               && The number of edges in $\mathbf{E}^{[y]}$ &\\
& $v_{i}^{[x]}$                                           && The $i$-th node in $\mathbf{V}^{[x]}$ &\\
& $l_{i}^{[x]}$                                           && The community membership of $v_{i}^{[x]}$ &\\
& $V_{\alpha}^{[x]}$                                      && The $\alpha$-th community in $\mathbf{V}^{[x]}$ &\\
\bottomrule
\end{tabularx}
\end{table}

\section{Method Based on Optimizing the Composite Modularity}\label{sec4}
In this section, we propose our method for the problem formulated in Section~\ref{sec3}. Our method follows the line of the modularity optimization method which is widely used for detecting communities in homogeneous single-relational networks. The idea is to define a quality function modularity for evaluating the goodness of community partitions, and then search a partition with a high modularity score. Inspired by this idea, we propose a new quality function --- the composite modularity for evaluating the goodness of a partition of a heterogeneous multi-relational network into communities, and develop a fast algorithm for optimizing the composite modularity. In the following, we first propose the composite modularity, and then present its optimization algorithm. For convenience, we list the major notations in Table~\ref{table1}.

\subsection{The Composite Modularity}\label{sec4.1}
\begin{figure}[!t]
\centering
\includegraphics[width=0.95\textwidth]{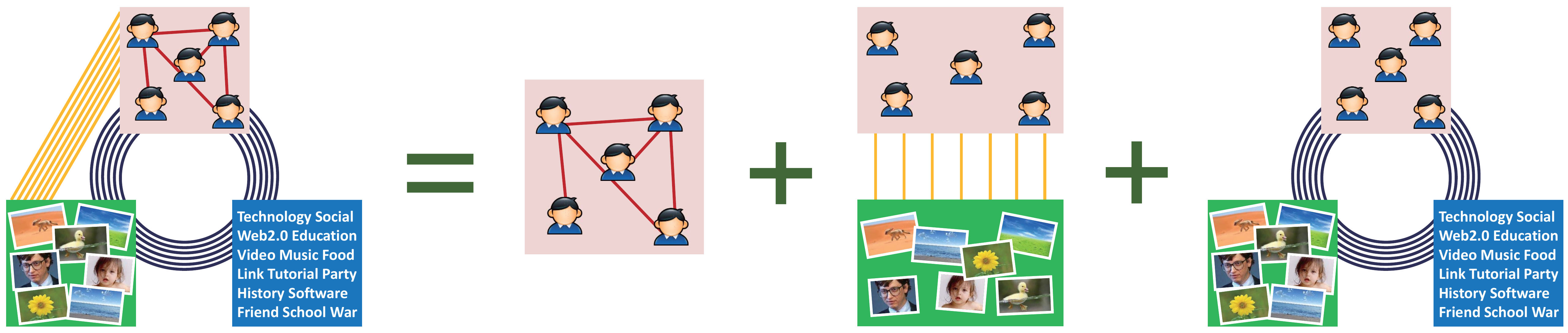}
\caption{\label{fig5} The Flickr heterogeneous multi-relational network can be regarded as an integration of a unipartite subnetwork, a bipartite subnetwork, and a tripartite subnetwork.}
\end{figure}

A heterogeneous multi-relational network $\mathbf{G} = (\mathbf{V}^{[1]} \cup \mathbf{V}^{[2]} \cup \cdot\!\cdot\!\cdot \cup \mathbf{V}^{[r]},\,\mathbf{E}^{[1]} \cup \mathbf{E}^{[2]} \cup \cdot\!\cdot\!\cdot \cup \mathbf{E}^{[s]})$ contains multiple types of nodes and edges. In another way, we can take $\mathbf{G}$ as an integration of multiple subnetworks. For example, we can take the Flickr heterogeneous multi-relational network as an integration of three subnetworks, as illustrated in Fig.~\ref{fig5}. Suppose $\mathbf{G}^{[y]}$ denotes the subnetwork which consists of $\mathbf{E}^{[y]}$ and the incident nodes. We can represent $\mathbf{G}$ as $\mathbf{G} = \mathbf{G}^{[1]} \cup \mathbf{G}^{[2]} \cup \cdots \cup \mathbf{G}^{[s]}$.

For each subnetwork, either a unipartite subnetwork or a k-partite subnetwork, researchers have proposed a modularity for evaluating partition of the related node sets \cite{NewmanModularityDefinition,MurataBipartiteModularity,MurataTriModularityWWW}. We can integrate these modularities and define a composite modularity for evaluating partition of all node sets as follows
\begin{eqnarray}
Q(\mathcal{L}) = \sum_{y=1}^{s} \frac{m^{[y]}}{m} Q^{[y]}(\mathcal{L}). \label{eq5}
\end{eqnarray}
Here $m^{[y]} = |\mathbf{E}^{[y]}|$ is the number of edges in $\mathbf{G}^{[y]}$, $m = \sum_{y=1}^{s} m^{[y]}$ is the total number of edges, and $Q^{[y]}$ is the modularity in $\mathbf{G}^{[y]}$.

For a unipartite subnetwork $\mathbf{G}^{[y]}$, $Q^{[y]}$ is the modularity proposed by Newman and Girvan \cite{NewmanModularityDefinition}. For convenience of description, we suppose all nodes of $\mathbf{V}^{[x]}$ are contained in $\mathbf{G}^{[y]}$, or mathematically, $\mathbf{G}^{[y]} = (\mathbf{V}^{[x]}, \mathbf{E}^{[y]})$. The partition on $\mathbf{V}^{[x]}$ is $\mathcal{L}^{[x]}=\{V_{\alpha}^{[x]}\,|\,\alpha=1,\cdot\!\cdot\!\cdot,c^{[x]}\}$ \footnote{In the case that not all nodes of $\mathbf{V}^{[x]}$ are contained in $\mathbf{G}^{[y]}$, for example, $\mathbf{G}^{[y]} = (\mathbf{V'}^{[x]}, \mathbf{E}^{[y]})$, where $\mathbf{V'}^{[x]}\subset\mathbf{V}^{[x]}$. Then the partition on $\mathbf{V'}^{[x]}$ is $\mathcal{L'}^{[x]} = \{V_{\alpha}^{[x]}\cap\mathbf{V'}^{[x]}\,|\,V_{\alpha}^{[x]}\cap\mathbf{V'}^{[x]}\neq\emptyset, \alpha=1,\cdot\!\cdot\!\cdot,c^{[x]}\}$.}, where $V_{\alpha}^{[x]}$ denotes the $\alpha$-th community in $\mathbf{V}^{[x]}$, and $c^{[x]}$ denotes the number of communities in $\mathbf{V}^{[x]}$. Then
\begin{eqnarray}
Q^{[y]}(\mathcal{L}) = Q^{[y]}(\mathcal{L}^{[x]}) = \sum_{{\alpha}=1}^{c^{[x]}} (e_{\alpha\alpha}^{[y]}-{a_{\alpha}^{[y]}}^2), \label{eq6}
\end{eqnarray}
where
\begin{eqnarray}
e_{\alpha\alpha}^{[y]} = \frac{1}{2m^{[y]}} \sum_{i=1}^{n^{[x]}}\sum_{j=1}^{n^{[x]}} A_{ij}^{[y]} \delta(l_{i}^{[x]},\alpha)\delta(l_{j}^{[x]},\alpha) \label{eq7}
\end{eqnarray}
is the fraction of subnetwork edges within community $V_{\alpha}^{[x]}$, and
\begin{eqnarray}
a_{\alpha}^{[y]} = \frac{1}{2m^{[y]}} \sum_{i=1}^{n^{[x]}}\sum_{j=1}^{n^{[x]}} A_{ij}^{[y]} \delta(l_{i}^{[x]},\alpha) \label{eq8}
\end{eqnarray}
is the fraction of subnetwork edges attached to community $V_{\alpha}^{[x]}$.


For a k-partite subnetwork $\mathbf{G}^{[y]}$, $Q^{[y]}$ is a variant k-partite modularity proposed by us previously \cite{MurataBipartiteModularity,MurataTriModularityWWW}. For convenience of description, we also suppose all nodes of $\mathbf{V}^{[x_1]}, \mathbf{V}^{[x_2]},\cdots,\mathbf{V}^{[x_k]}$ are contained in $\mathbf{G}^{[y]}$, or mathematically, $\mathbf{G}^{[y]} = (\mathbf{V}^{[x_1]}\cup\mathbf{V}^{[x_2]}\cup\cdots\cup\mathbf{V}^{[x_k]}, \mathbf{E}^{[y]})$. The partition on $\mathbf{V}^{[x_1]}, \mathbf{V}^{[x_2]},\cdots,\mathbf{V}^{[x_k]}$ is $\mathcal{L}^{[x_1]}\cup\mathcal{L}^{[x_2]}\cup\cdots\cup\mathcal{L}^{[x_k]}=\{V_{\alpha_{1}}^{[x_1]}\,|\,\alpha_{1}=1,\cdots,c^{[x_1]}\} \cup \{V_{\alpha_{2}}^{[x_2]}\,|\,\alpha_{2}=1,\cdots,c^{[x_2]}\} \cup \cdots \cup \{V_{\alpha_{k}}^{[x_k]}\,|\,\alpha_{k}=1,\cdots,c^{[x_k]}\}$ \footnote{In the case that not all nodes of $\mathbf{V}^{[x_1]}, \mathbf{V}^{[x_2]},\cdots,\mathbf{V}^{[x_k]}$ are contained in $\mathbf{G}^{[y]}$, please refer to the solution for a similar case in the unipartite subnetwork.}. Then,
\begin{eqnarray}
Q^{[y]}(\mathcal{L}) &=& Q^{[y]}(\mathcal{L}^{[x_1]}\cup\mathcal{L}^{[x_2]}\cup\cdot\!\cdot\!\cdot\cup\mathcal{L}^{[x_k]}) \\
&=& \frac{1}{k} \sum_{t=1}^{k} \sum_{\alpha_t=1}^{c^{[x_t]}} (e_{{\alpha_1}\cdot\cdot\cdot{\alpha_k}}^{[y]} - a_{\alpha_1}^{[x_1,y]} \cdots a_{\alpha_k}^{[x_k,y]})
\end{eqnarray}
where
\begin{eqnarray}
{\alpha_1},\cdots,{\alpha_{t-1}},{\alpha_{t+1}},\cdots,{\alpha_k} ={\argmax_{{\tilde{\alpha_1}},\cdots,{\tilde{\alpha_{t-1}}},{\tilde{\alpha_{t+1}}},\cdots,{\tilde{\alpha_k}}}} e_{{\tilde{\alpha_1}}\cdot\!\cdot\!\cdot{\tilde{\alpha_{t-1}}},{\alpha_{t}}{\tilde{\alpha_{t+1}}}\cdot\!\cdot\!\cdot{\tilde{\alpha_k}}}^{[y]}
\end{eqnarray}
indicate the indices of the communities which have the largest number of subnetwork edges to community $V_{\alpha_t}^{[x_t]}$,
\begin{eqnarray}
e_{{\alpha_1}\cdot\!\cdot\!\cdot{\alpha_k}}^{[y]} = \frac{1}{m^{[y]}} \sum_{i_1=1}^{n^{[x_1]}} \cdot\!\cdot\!\cdot \sum_{i_k=1}^{n^{[x_k]}} A_{i_1\cdots i_k}^{[y]} \delta(l_{i_1}^{[x_1]},\alpha_1) \cdot\!\cdot\!\cdot \delta(l_{i_k}^{[x_k]},\alpha_k) \label{eq9}
\end{eqnarray}
is the fraction of subnetwork edges between community $V_{\alpha_1}^{[x_1]}, V_{\alpha_2}^{[x_2]}, \cdots, V_{\alpha_k}^{[x_k]}$, and
\begin{eqnarray}
a_{\alpha_t}^{[x_t,y]} = \frac{1}{m^{[y]}} \sum_{i_1=1}^{n^{[x_1]}} \cdot\!\cdot\!\cdot \sum_{i_k=1}^{n^{[x_k]}} A_{i_1\cdots i_k}^{[y]} \delta(l_{i_t}^{[x_t]},\alpha_t) \label{eq10}
\end{eqnarray}
is the fraction of subnetwork edges attached to community $V_{\alpha_t}^{[x_t]}$.

In general $Q$ ranges in $[-1,1]$, as each $Q^{[y]}$ ranges in $[-1,1]$. Note that the composite modularity is consistent with Newman and Girvan's modularity \cite{GirvanNewmanCommuityDefinition} and our k-partite modularity \cite{MurataBipartiteModularity,MurataTriModularityWWW} --- If the heterogeneous multi-relational network reduces to a unipartite network/k-partite network, the composite modularity recovers Newman and Girvan's modularity/our k-partite modularity.

In the above definition of the composite modularity, the component modularities are weighted by the fraction of edges of the subnetwork. An underlying assumption of this weighting strategy is that edges are treated equally, no matter their types. This is a natural assumption in the case where we have no background knowledge about the network. However, in the case that some type of edges account for a dominant proportion of the total number of edges, or some type of edges contain much noise and should be attached less weight, the above definition may be limited in detecting the communities. In these cases, we should use other weighting strategies based on some a priori knowledge of the network. For example, we can define the composite modularity as
\begin{eqnarray}
Q(\mathcal{L}) = \sum_{y=1}^{s} \frac{1}{w^{[y]}} Q^{[y]}(\mathcal{L}), \label{eq11}
\end{eqnarray}
where $w^{[y]}$ represents the relative importance of the $y$-th type edges, and is specified by the experimenter.

\subsection{Optimization Algorithm}\label{sec4.2}
Optimizing the composite modularity is NP-hard \cite{BrandesModularityNPCompleteness}. We develop an efficient heuristic algorithm for practical use. Our algorithm is based on the Louvain algorithm, originally designed by Blondel et al. for optimizing Newman and Girvan's modularity \cite{BlondelFastUnfolding}. In the following, we first describe how to optimize the composite modularity by the Louvain algorithm, and then discuss how this algorithm can be improved for practical application.

Louvain algorithm contains two phases which are performed iteratively until maximal $Q$ is reached. In the first phase, each node is initially assigned to a community of itself. Then, for each node $v_{i}^{[x]}$ ($x \in \{1,2,\cdot\!\cdot\!\cdot,r\}$, $i \in \{1,2,\cdot\!\cdot\!\cdot,n^{[x]}\}$), the gains in $Q$ that would result from moving $v_{i}^{[x]}$ to the community of each of its counterparts ($v_{i}^{[x]}$'s 1-hop and 2-hop neighbors which are of the same node type as $v_{i}^{[x]}$) are calculated, and $v_{i}^{[x]}$ is moved to the community for which the maximum positive gain in $Q$ is attained. If no positive gain is possible, $v_{i}^{[x]}$ stays in its original community. This subprocess is repeated sequentially for all nodes until no individual move will result in a gain, marking the end of the first phase. In the second phase, a new network is built whose nodes are the communities in the first phase. The weights of the edges between the new nodes are calculated as the sum of the weights of the edges between nodes in the corresponding communities. Once this second phase is completed, the two phases are repeated iteratively until there are no changes.

The efficiency of the first phase relies on calculation of the gains in $Q$ resulting from moving $v_{i}^{[x]}$ to the community of each of its counterparts. This operation requires a computational complexity of $O({\bar{d}}^{2k_{max}})$ in the worst case, where $\bar{d}$ is the average node degree and $k_{max}$ is the highest partite number for the k-partite subnetworks. Thus the complexity of the first phase is $O({\bar{d}}^{2k_{max}})$. The complexity of the second phase is $O(n)$. In practice, the number of iterations of the two phases is generally small \cite{BlondelFastUnfolding}. Consequently, the total computational complexity of Louvain algorithm for optimizing the composite modularity is near $O({\bar{d}}^{2k_{max}}{\cdot}n)$. This complexity may prohibit practical implementations for large-scale networks which contain millions of nodes.

\begin{algorithm}[!t]
\KwIn{The community partitions for each subnetwork (each community has a global unique ID)}
\KwOut{Sets of nodes which are assigned to the same community by every related partition}
\Begin{
    Iterate the partitions for each subnetwork and build a list for each node $v_{i}^{[x]}$. The list records the IDs of the communities to which $v_{i}^{[x]}$ belongs (referred as $v_{i}^{[x]}$'s community trace).\;
    \BlankLine
    \tcp{Nodes assigned to the same community by every related partition should have the same community trace, and vice versa.}
    Examine the above records and build a map, where a key is a community trace and its value is the nodes which have this community trace.\;
    \BlankLine
    Then each value represents a set of nodes which must be assigned together. Return these sets as output.\;
}
\caption{Find the \textit{must-be-assigned-together} constraints}\label{alg1}
\end{algorithm}

To speed up the algorithm, we propose a Louvain-C algorithm with the strategy of \textit{divide and rule}. Given a heterogeneous multi-relational network which is an integration of multiple subnetworks, Louvain-C algorithm consists of the following three steps.
\begin{itemize}
\item[(1)] Detect communities in each subnetwork seperately;
\item[(2)] Combine the partitions obtained in each subnetwork and derive some constraints;
\item[(3)] Optimize the composite modularity under these constraints.
\end{itemize}
As for step (1), we can use modularity and k-partite modularity optimization methods to detect communities in each subnetwork. Other existing methods (see Section~\ref{sec2}) can also be used. Since a node can be involved in multiple subnetworks, it may be assigned to different communities by partitions in different subnetworks. Thus, we combine these partitions and derive some consistent constraints in step (2). Nodes $v_{i_1}^{[x]}$, $v_{i_2}^{[x]}$, $\cdot\!\cdot\!\cdot$, $v_{i_k}^{[x]}$ form a \textit{must-be-assigned-together} constraint if they are assigned to the same community by every related partition. Algorithm~\ref{alg1} shows the procedure of finding such constraints. The computational complexity of Algorithm~\ref{alg1} is $O(n{\cdot}logn)$. In step (3), we first build a new network based on these constraints, in a similar way as the second phase of Louvain algorithm. In other words, a node in the new network corresponds to a group of nodes that must be assigned together and weights between the new nodes are recalculated. Then, we optimize the composite modularity in this new network. Note that the size of the new network is much smaller than the original one. On the other hand, building the new network has little impact on the optimization result, since we just aggregate the nodes that ``must be assigned together''. As a result, Louvain-C algorithm can dramatically reduce the computational time without compromising the accuracy. Please see the next section for more information about comparison of the two algorithms.

\section{Experiments}\label{sec5}
So far we have proposed the composite modularity optimization method for detecting communities in heterogeneous multi-relational networks. In this section, we present experiments for testing performance of our method.

\subsection{Comparison with Other Methods in Synthetic Networks}\label{sec5.1}
First, we compare our method with others in synthetic networks. The basic scheme is as follows. 1) We generate a sequence of synthetic heterogeneous multi-relational networks with planted communities. 2) Applying various methods to these networks (the planted partition is hidden at this time), we test which method can detect most of the planted communities. Such kind of testing is widely used by other researchers in the community detection field \cite{DanonCompareCommunityAlgo,GirvanNewmanCommuityDefinition,LancichinettiLFRBenchmark,LancichinettiDirectedLFRBenchmark,LancichinettiCommunityAlgoAnalysis}.

\begin{figure}[!t]
\centering
\includegraphics[width=0.60\textwidth]{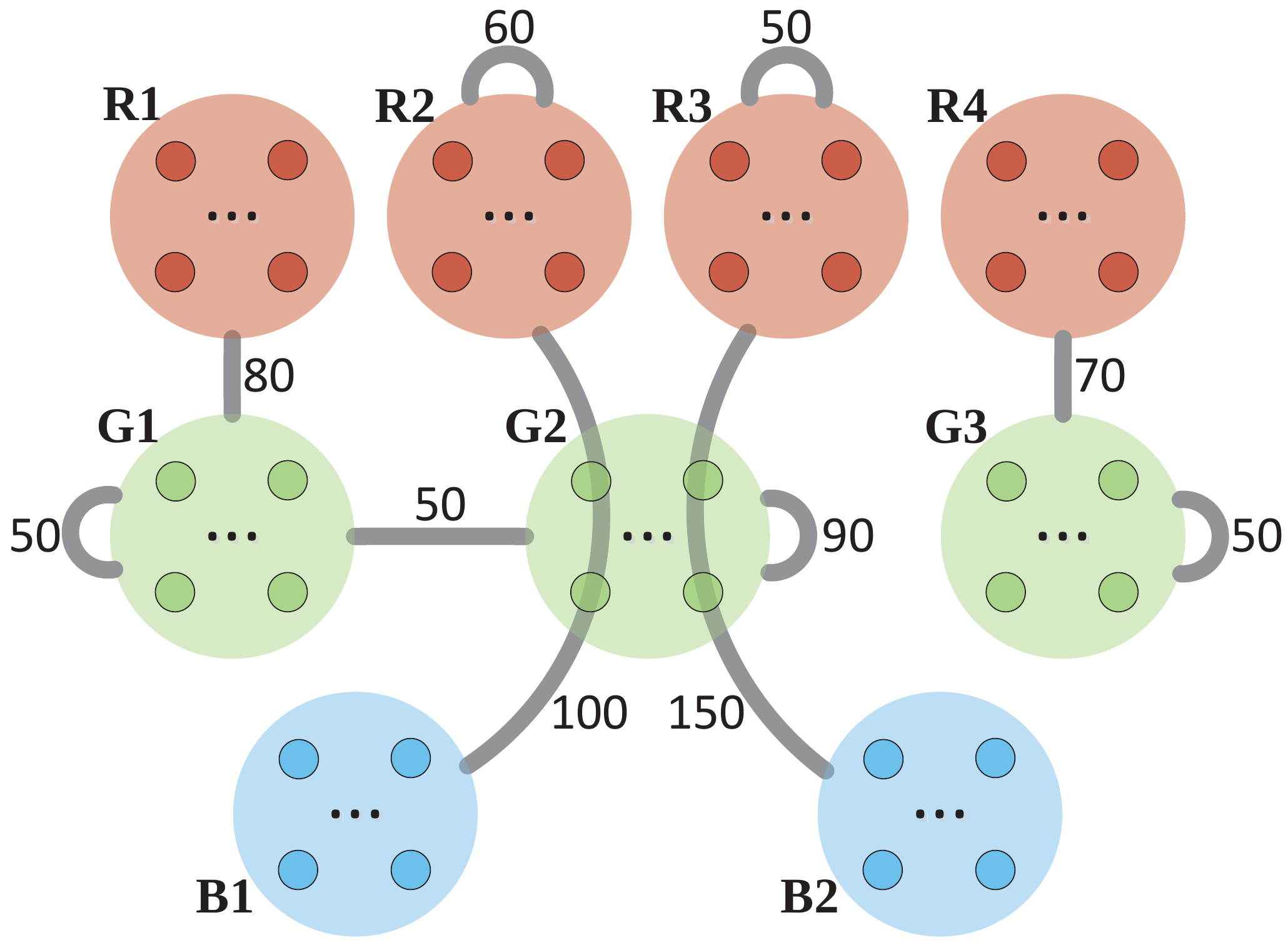}
\caption{\label{fig6} The link patterns of the communities in the synthetic network. The numbers indicate the numbers of edges.}
\end{figure}

As shown in Fig.~\ref{fig6}, our synthetic network model contains three types of nodes (the red, green, and blue nodes), and four types of edges (the edges between red nodes, the edges between green nodes, the edges between red and green nodes, and the hyper-edges between red, green, and blue nodes). Red nodes are organized into four communities, each containing 15 nodes. Green nodes are organized into three communities, each containing 20 nodes. Blue nodes are organized into two communities, each containing 25 nodes. From Fig.~\ref{fig6} we can see that each community has its own representative link pattern (the number of edges are shown in the figure). For example, the link pattern of community G1 is that its nodes all densely connect to nodes in G1 (edge density = $\frac{50}{\text{Size(G1)}\times[\text{Size(G1)}-1]/2}=0.2632$), to nodes in R1 (edge density = $\frac{80}{\text{Size(G1)}\times\text{Size(R1)}}=0.2667$), and to nodes in G2 (edge density = $\frac{50}{\text{Size(G1)}\times\text{Size(G1)}}= 0.1250$). Based on these link patterns, we generated a total of 750 edges. Then, we add noise edges randomly. The noise rate increases from $0\%$ to $200\%$ (thus the total number of edges ranges from 750 to 2,250), so that the planted communities are more and more difficult to be detected.

To evaluate a method, we apply it to the network and calculate the similarity between the obtained partition and the planted partition. The more similar the two partitions, the better the method. We adopted the regularly used \textit{normalized mutual information} (NMI) \cite{FredNMI,DanonCompareCommunityAlgo} to quantify the similarity between two partitions $\mathcal{L}_1$ and $\mathcal{L}_2$. If $\mathcal{L}_1$ and $\mathcal{L}_2$ match completely, we have a maximum NMI value of 1, whereas if $\mathcal{L}_1$ and $\mathcal{L}_2$ are totally independent of one another, we have a minimum value of 0.

We compare our method with the following four methods, which cover the state-of-the-art techniques.
\begin{itemize}
\item \textit{NaiveSimp}: Simplify the heterogeneous multi-relational network to a single-relational network and detect communities (the results are based on the best performance obtained in the single-relational network for each edge type).
\item \textit{Trans-CN} (A modification of the method proposed in Ref.\,\cite{PopesculClusteringBasedOnSimilarity}): Transform the heterogeneous multi-relational network to weighted unipartite networks for each node type based on the number of common neighbors. Then detect communities in each unipartitie network separately.
\item \textit{Trans-JD} (A modification of the method proposed in Ref.\,\cite{PopesculClusteringBasedOnSimilarity}): Transform the heterogeneous multi-relational network to weighted unipartite networks for each node type based on the Jaccard index. Then detect communities in each unipartitie network separately.
\item \textit{MetaFac} \cite{LinMetaFac}: Detect communities based on tensor factorization. This method assumes that nodes of different types have the same number of communities, and requires this number as an input. We set this number to four, the number of red communities.
\end{itemize}

\begin{figure}[!t]
\centering
\subfigure[][]{\label{fig7a}
\includegraphics[width=0.47\textwidth]{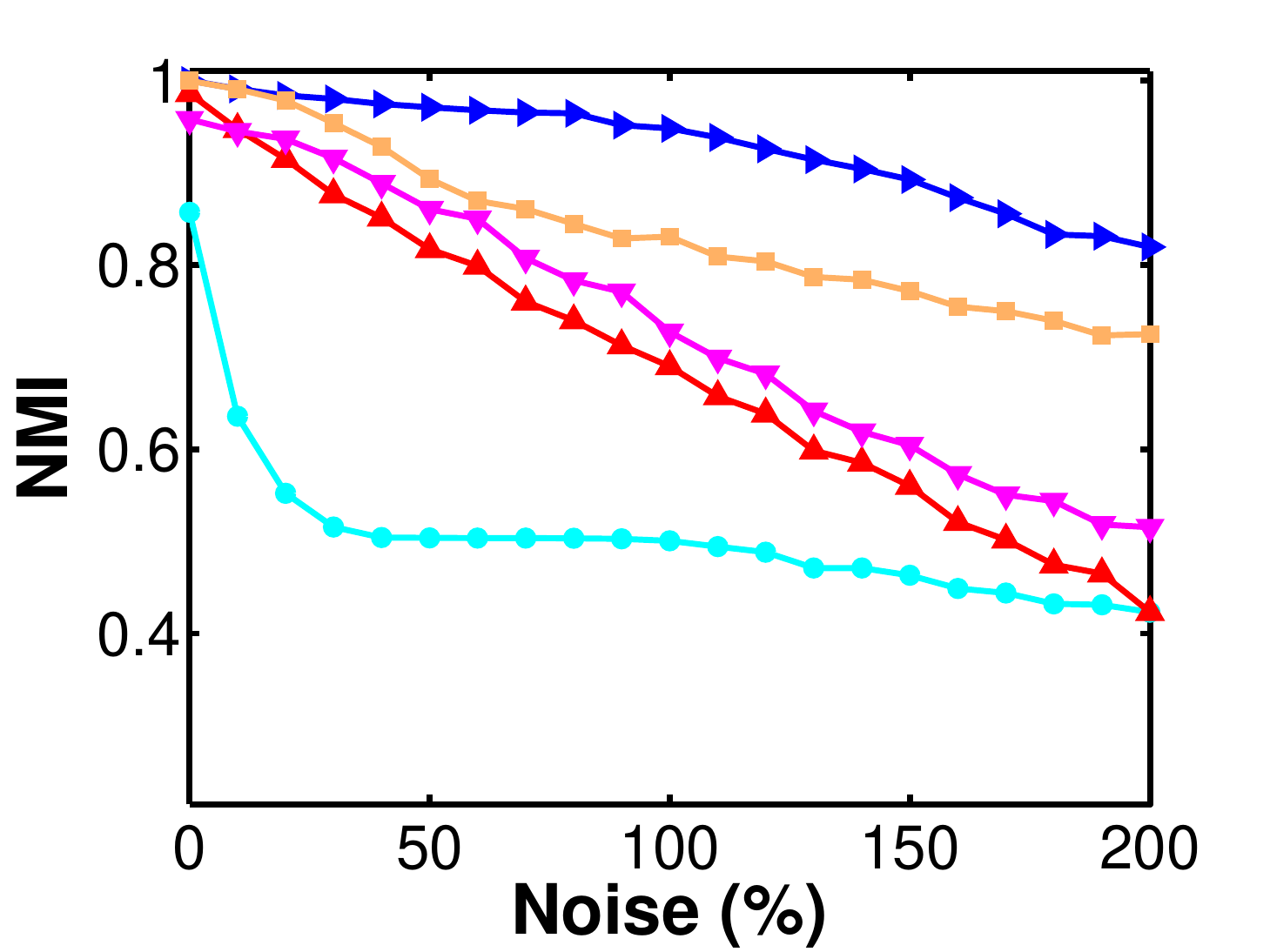}
}\\
\subfigure[][]{\label{fig7b}
\includegraphics[width=0.47\textwidth]{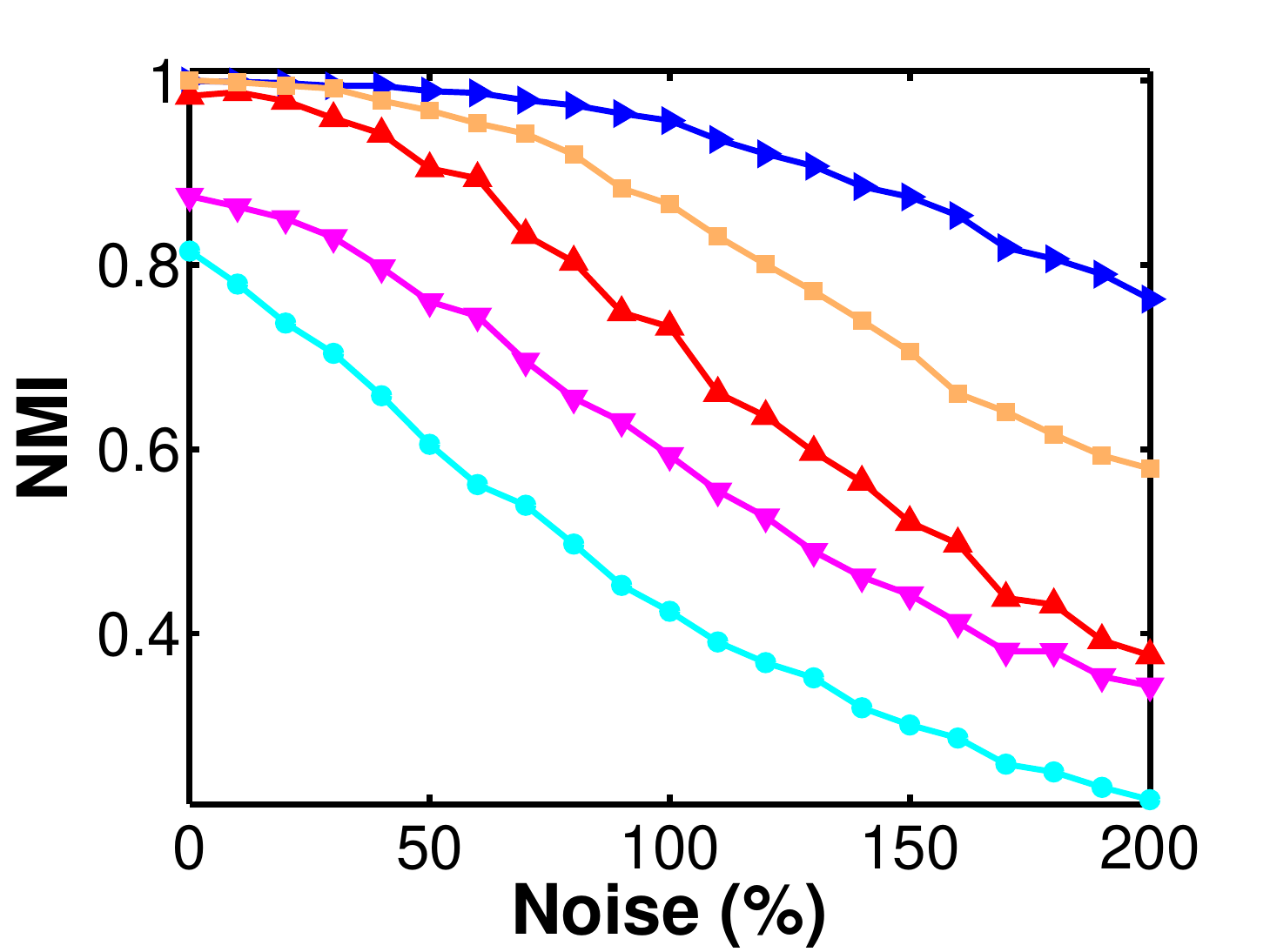}
}\\
\subfigure[][]{\label{fig7c}
\includegraphics[width=0.47\textwidth]{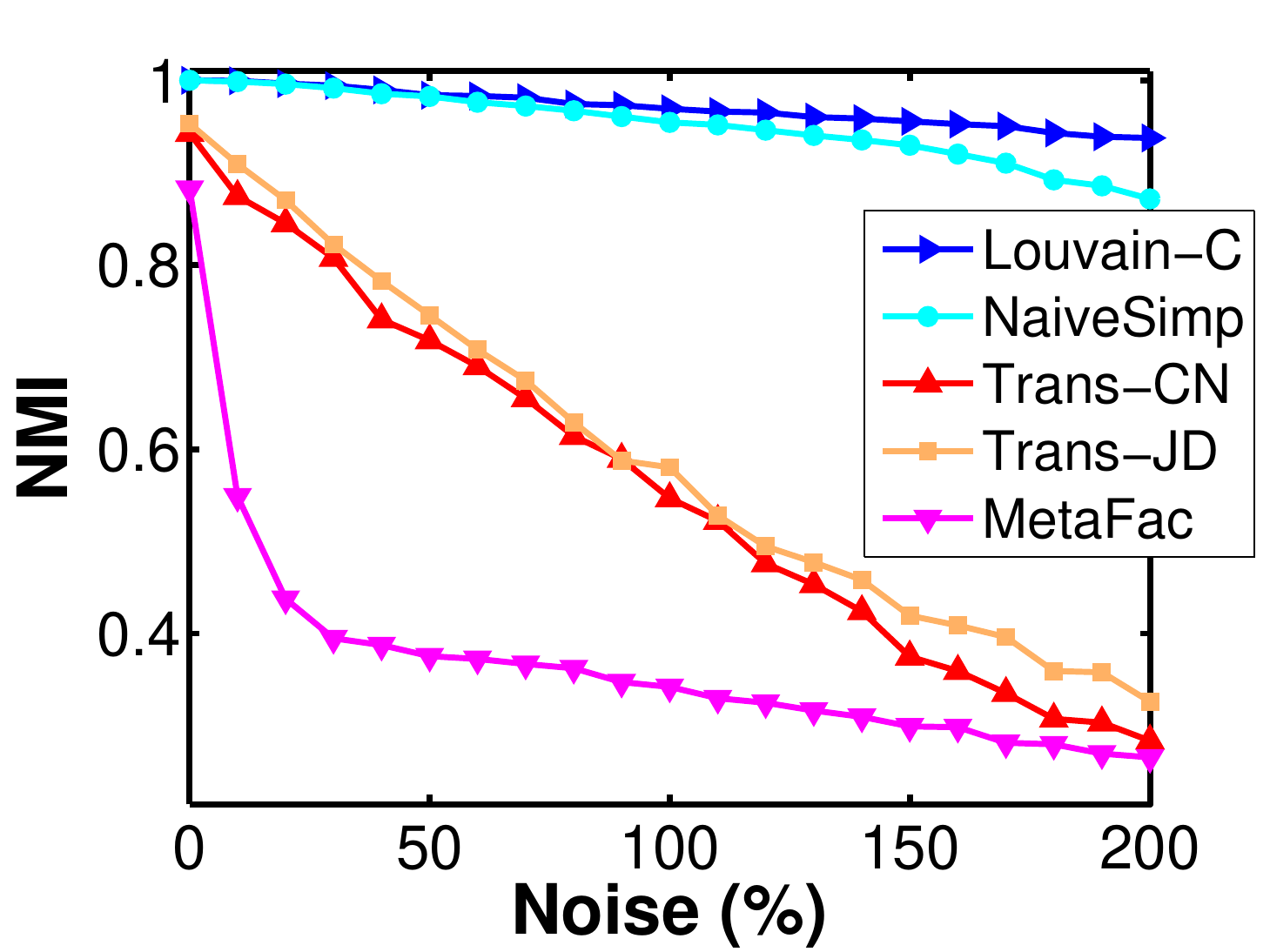}
}
\caption{\label{fig7} The NMI values achieved by different methods in the synthetic networks for (a) the red node set, (b) the green node set, and (c) the blue node set.}
\end{figure}

The results are shown in Fig.~\ref{fig7}. We can find that Louvain-C algorithm outperforms the others by a large margin. It successfully detected the planted communities in the red, green, and blue node sets when the noise is $0\%$. When the noise goes up to $200\%$, the NMI values are still higher than 0.75. As for other methods, none of them can detect the planted communities with $100\%$ accuracy even when the noise is $0\%$. In specific, the inferiority of Trans-JD and Trans-CN is due to the fact that we cannot rely on local measures (common neighbor index and Jaccard Index) to accurately calculating the similarity between nodes. Moreover, as the noise increases, the transferred unipartite networks of the two methods become so dense that almost all pairs of nodes are connected. Thus, it makes it increasingly difficult to detect communities. As a result, the performances of Trans-JD and Trans-CN plummet dramatically as the noise increases. NaiveSimp has good performance in the blue node set, but does not work well in the red and green node sets. This is because the information contained in the simplified single-relational network is sometimes incomplete. For example, in the unipartite subnetwork of green nodes there are dense edges both within and between the community G1 and G2. Thus, NaiveSimp failed to separate them and took them as a single community. The performance of MetaFac is also not so remarkable, especially in the blue node set. The reason is that this method assumes that nodes of different types have the same number of communities. However, red, green, and blue nodes have different number of planted communities. In summary, this experiment shows that our method is better than the state-of-the-art techniques in detecting the planted communities.

\subsection{Scalability}\label{sec5.2}
In this section, we test the scalability of our method. We gradually increase the size of the synthetic network and compare Louvain-C and Louvain algorithm. Since these two algorithms are for optimizing the composite modularity, we are interested in comparing their performance in terms of running time and the composite modularity value.

\begin{figure}[!t]
\centering
\setlength{\abovecaptionskip}{1.5pt}
\subfigure[][]{\label{fig8a}
\includegraphics[width=0.47\textwidth]{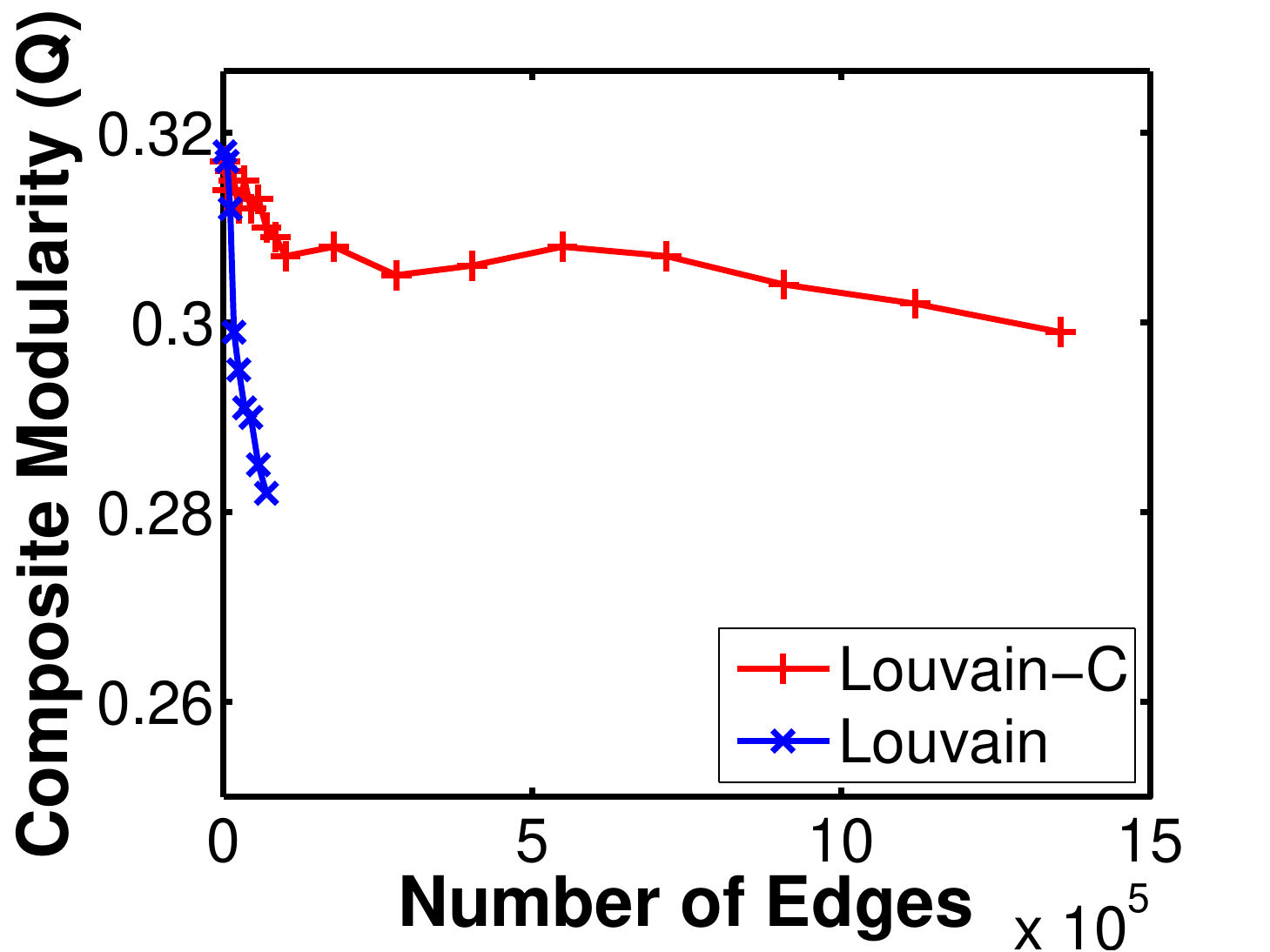}
}
\subfigure[][]{\label{fig8b}
\includegraphics[width=0.47\textwidth]{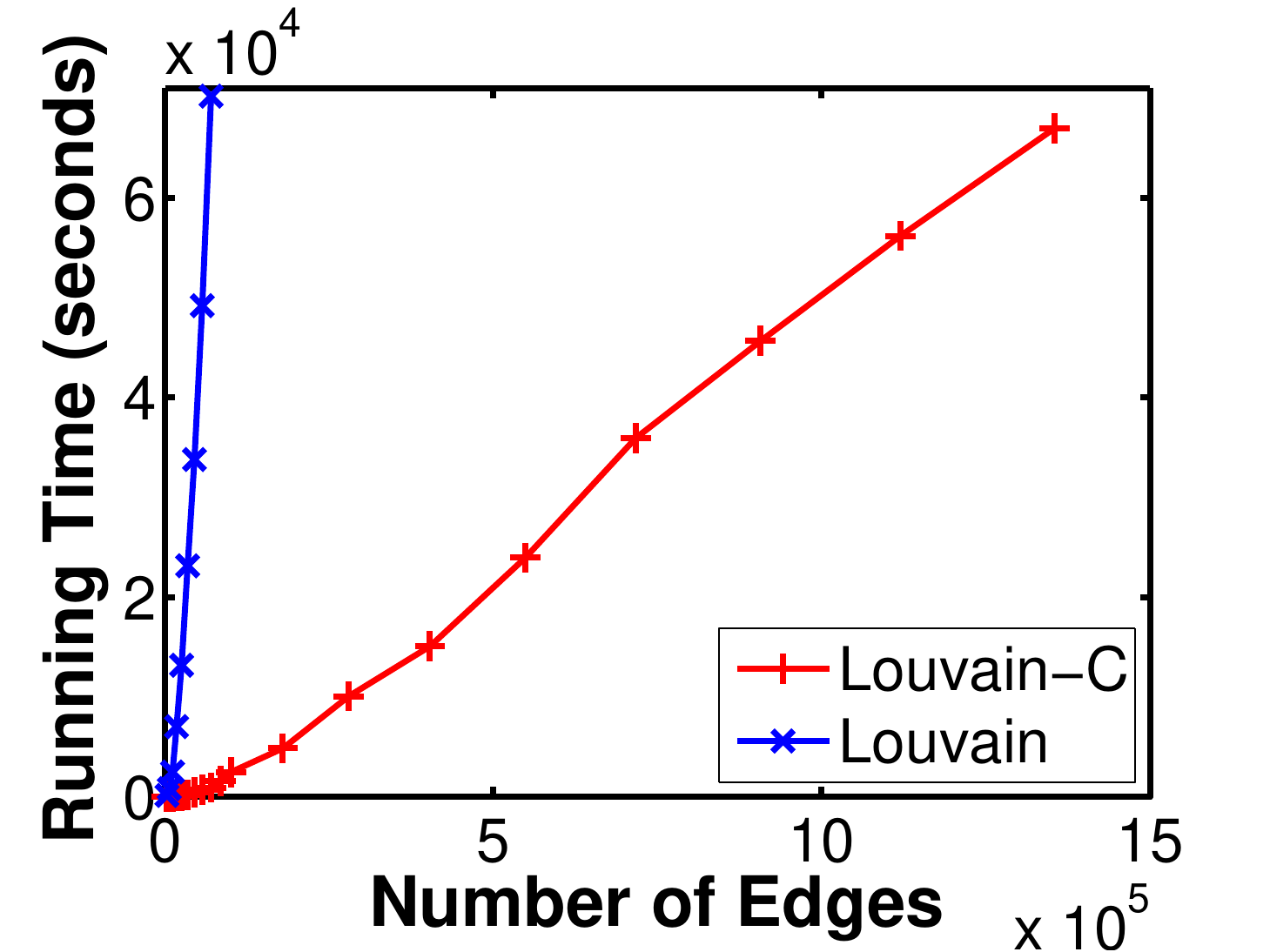}
}
\caption{\label{fig8} (a) The composite modularity values achieved by Louvain-C and Louvain algorithm. (b) The running time of Louvain-C and Louvain algorithm. The algorithms are built in Python 3.2, and run on a PC equipped with an Intel Core i7-2600 CPU at 3.40 GHz and 32GB physical memory.}
\end{figure}

Fig.~\ref{fig8a} shows the composite modularity values achieved by the two algorithms. We can see that when the network size is small, Louvain algorithm is competitive with Louvain-C algorithm, but as the network size increases, Louvain-C algorithm achieved higher values than Louvain algorithm. The reason is as follows. As the network size increases, the search space increases dramatically. As a result, Louvain algorithm converged to local maxima. However, the search space can be greatly reduced, given the must-be-assigned-together constraints. Thus, Louvain-C algorithm produced better results.

Fig.~\ref{fig8b} shows that Louvain-C algorithm is much faster than Louvain algorithm. For example, when the network size is 70,000 edges, Louvain algorithm took 70,199.6 seconds while Louvain-C algorithm took only 911.1 seconds, a reduction of around 77 times. Thus, Louvain-C algorithm can search better solutions than Louvain algorithm in much less time.

\begin{figure}[!t]
\centering
\includegraphics[width=0.98\textwidth]{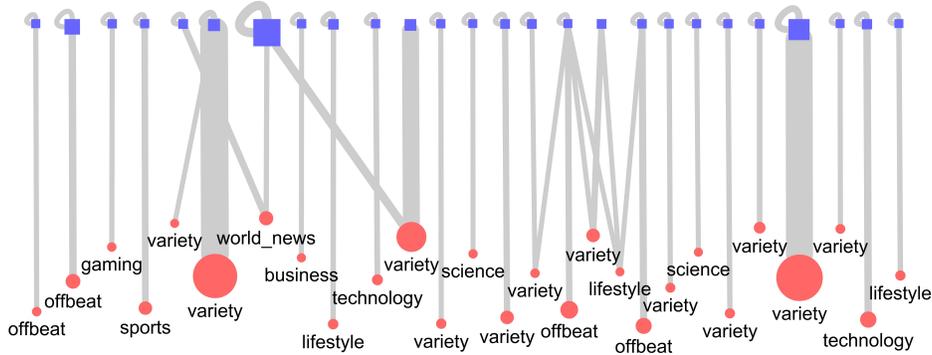}
\caption{\label{fig9} Visualization of the Digg network at the community level (small communities of less than 10 nodes, and sparse connections with density less than 0.05 are omitted). The story and user community are colored in red and blue, respectively. The label of a story community denotes its dominant topic (\textit{variety} means that there is no dominant topic).}
\end{figure}

\subsection{Application in Digg Network}\label{sec5.3}
Finally, we apply Louvain-C algorithm to a real-world Digg network. Digg is a social news website. Digg users can submit web content as a story and other users can vote the story by ``digging" it. In addition, users can add other users as their friends. We collected a subset of the stories submitted during Oct 8-15, 2010. Then we constructed a heterogeneous multi-relational network, which contains 1,191 users, 3,610 stories, 7,318 edges representing the friendship between users, and 20,941 edges representing the digging relationship between users and stories.

Louvain-C algorithm detected 90 user communities and 88 story communities in the Digg network. The composite modularity is 0.4203. We can check the reasonability of our results by looking at the topic associated with each story. In total there are 10 predesignated topics (business, entertainment, gaming, lifestyle, offbeat, politics, science, sports, technology, and world news), and Digg users can assign one of the topics to a story during submission. We found that many story communities have dominant topics. This means that a great many users are interested in stories with determined topics. This makes sense because people always focus on limited stories based on their personal interests.

Fig.~\ref{fig9} visualizes the Digg network at the community level. From this figure, we can gain knowledge about which user community is interested in which story community. We find that most user communities have only one corresponding story community, while only a few have several corresponding story communities. In addition, we can predict the interests of a user community based on the topic of its corresponding story community. This is a significant advantage of our method over NaiveSimp, Trans-CN, and Trans-JD, since they only consider one type of nodes at a time and cannot gain such information that are hidden between different types of nodes.

\begin{figure}[!t]
\centering
\setlength{\abovecaptionskip}{1pt}
\includegraphics[width=0.66\textwidth]{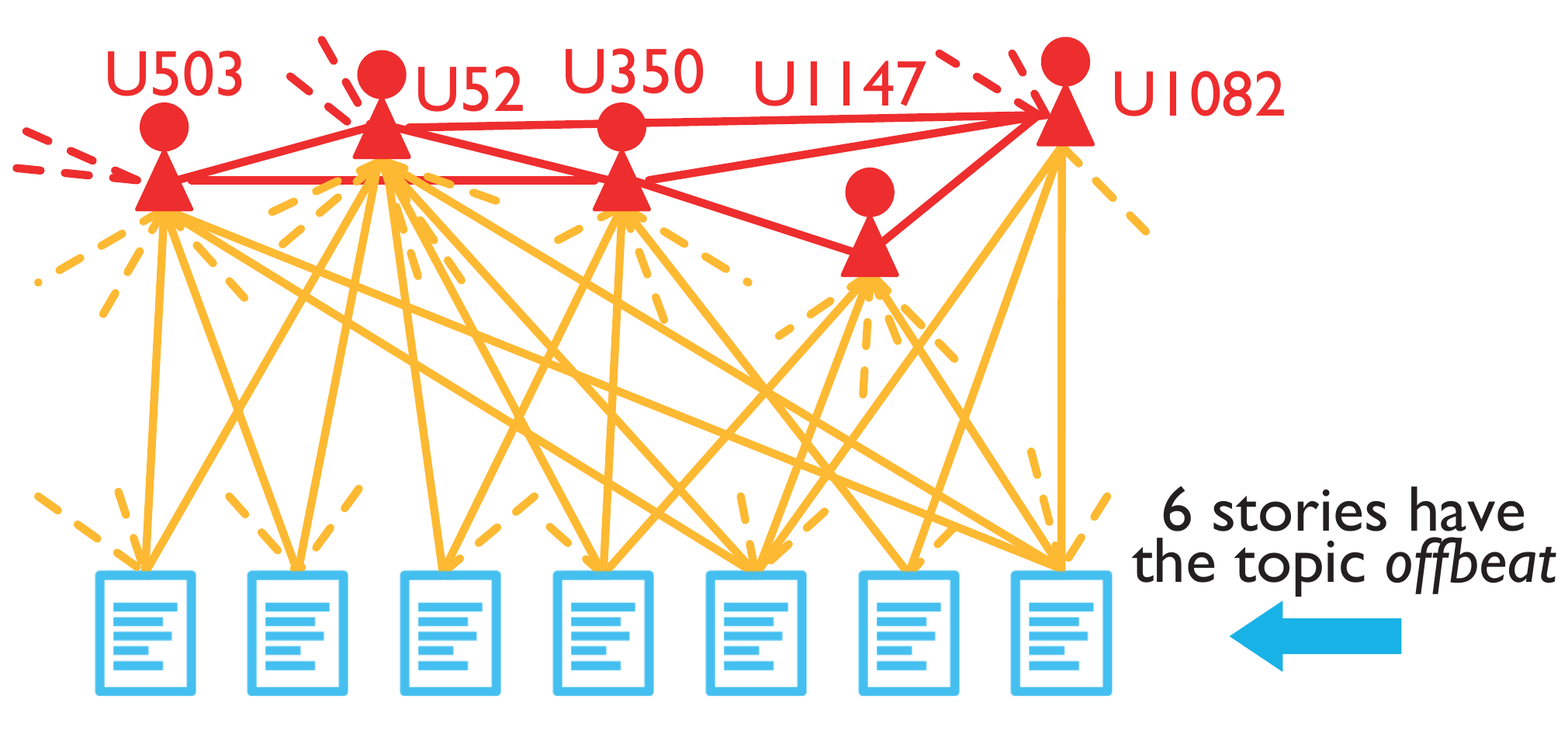}
\caption{\label{fig10} A user community and the corresponding story community.}
\end{figure}

To better understand how one can use our method for analysis, we focus on a user community and its corresponding story community, as shown in Fig.~\ref{fig10}. The user community contains 5 members who are densely connected to each other. The story community contains 7 members. Connections between the user community and the story community are dense. Obviously, the members of the user/story community have the similar link patterns. Thus, both communities are meaningful. Moreover, 6 members of the story community have the same topic \textit{offbeat}. Based on this, we can predict that these 5 users are especially interested in offbeat events. Mining such information is useful in targeted advertising.

Note that in Fig.~\ref{fig9} there are two user communities which have only sparse connections in between them. The identification of them is based on their link patterns with the story nodes. Clearly, the two communities cannot be detected by NaiveSimp, Trans-CN, or Trans-JD. In summary, this experiment shows that our method detected reasonable communities that agree with the natural human intuition and perception.

\section{Conclusion}\label{sec6}
Previous community detection methods overwhelmingly focused on the homogeneous single-relational network which contain only one type of nodes and edges. However, Many real-world systems are naturally described as heterogeneous multi-relational networks which contain multiple types of nodes and edges. In this paper, we proposed the first modularity measure --- the composite modularity for evaluating partitions of a heterogeneous multi-relational network into communities. The key idea of our composite modularity is to decompose a heterogeneous multi-relational network into multiple subnetworks, and integrate the modularities in each subnetworks. Compared to modularity and its variants for single-relational networks, our composite modularity can effectively utilize multi-faceted information in a heterogeneous multi-relational network and comprehensively evaluate a partition. We developed a fast algorithm for optimizing the composite modularity and detecting the communities in heterogeneous multi-relational networks. In short, our composite modularity optimization method has the following advantages:
\begin{itemize}
\itemsep=-0.6pt
\item It is consistent with the modularity optimization method for detecting communities in homogeneous single-relational networks;
\item It is applicable to networks of general structure, which may even contain hyper-edges;
\item It is parameter free and can automatically detect communities without any a priori knowledge such as the number of communities;
\item It is fast and scalable to large-scale networks.
\end{itemize}

A notable drawback of modularity optimization is the resolution limit, which refers to the incapability of detecting small communities in large-scale networks \cite{FortunatoResolutionLimit,GoodModularityDegeneracy,lancichinettiModularityLimit}. Although it has not been shown, we expect that the composite modularity optimization method introduced in this work has a similar resolution limit. The resolution limit is attributed to the globalization of the null model, which assumes that any node can connect to any others. Thus, redefining the null model and restricting such globalization are a possible way to solve the resolution limit. This is left for our future work.

\section*{Acknowledgements}
We gratefully thank Mr.\ Pen-Lin Chang for collecting the Digg data.

\bibliographystyle{ws-acs}
\bibliography{myRef29}

\end{document}